\documentclass[12pt,aps,prd,preprint,tightenlines,superscriptaddress,showpacs,nofootinbib]
{revtex4-1}
\usepackage{epsfig}
\usepackage{epstopdf}
\usepackage{amssymb,amsmath}

\newcommand{\bea}{\begin{eqnarray}}
\newcommand{\eea}{\end{eqnarray}}

 \makeatletter
\def\alt{\mathrel{\mathpalette\gl@align<}}
\def\agt{\mathrel{\mathpalette\gl@align>}}
\def\gl@align#1#2{\lower.6ex\vbox{\baselineskip\z@skip\lineskip\z@
\ialign{$\m@th#1\hfil##\hfil$\crcr#2\crcr\sim\crcr}}} \makeatother

\begin{document}
%\begin{flushright}
%Preprint Number
%\end{flushright}
%
\vspace*{1.0cm}

\begin{center}
\baselineskip 20pt 
{\Large\bf 
Non-minimal quartic inflation \\
in classically conformal U(1)$_X$ extended Standard Model 
}
\vspace{1cm}

{\large 
Satsuki Oda$^{~a, b}$, Nobuchika Okada$^{~c}$, Digesh Raut$^{~c}$, \\ 
and Dai-suke Takahashi$^{~a, b}$
}
\vspace{.5cm}

{
\baselineskip 20pt \it
$^a$Okinawa Institute of Science and Technology Graduate University (OIST), \\
Onna, Okinawa 904-0495, Japan  \\ 
$^b$Research Institute, Meio University, Nago, Okinawa 905-8585, Japan\\
$^c$Department of Physics and Astronomy, University of Alabama, \\
Tuscaloosa, Alabama 35487, USA \\
} 

\vspace{.5cm}

\vspace{1.5cm} {\bf Abstract}
\end{center}

We propose quartic inflation with non-minimal gravitational coupling
   in the context of the classically conformal U(1)$_X$ extension of the Standard Model (SM). 
In this model, the U(1)$_X$ gauge symmetry is radiatively broken through the Coleman-Weinberg mechanism,  
   by which the U(1)$_X$ gauge boson ($Z^\prime$ boson) 
   and the right-handed Majorana neutrinos acquire their masses. 
We consider their masses in the range of ${\cal O}$(10 GeV)$-{\cal O}$(10 TeV), 
   which are accessible to high energy collider experiments. 
The radiative U(1)$_X$ gauge symmetry breaking also generates a negative mass squared for the SM Higgs doublet, 
   and the electroweak symmetry breaking occurs subsequently.
We identify the U(1)$_X$ Higgs field with inflaton and calculate the inflationary predictions. 
Due to the Coleman-Weinberg mechanism, the inflaton quartic coupling during inflation, 
   which determines the inflationary predictions, is correlated to the U(1)$_X$ gauge coupling. 
With this correlation, we investigate complementarities between the inflationary predictions and 
  the current constraint from the $Z^\prime$ boson resonance search at the LHC Run-2  
  as well as the prospect of the search for the $Z^\prime$ boson and the right-handed neutrinos 
  at the future collider experiments.

\thispagestyle{empty}

%\bigskip
\newpage

\addtocounter{page}{-1}

%%%%%%%%%%%%%%%%%%%%%%%%%%
%\baselineskip 36pt
% Main body
%%%%%%%%%%%%%%%%%%%%%%%%%%
\baselineskip 18pt
%%%%%%%%%%%%%%%%%%%%%%%%%%

%%%%%%%%%%%%%%%%%%%%%%%%
\section{Introduction} 
%%%%%%%%%%%%%%%%%%%%%%%%
\label{sec:1}

Cosmological inflation \cite{inflation} 
   provides not only solutions to problems in the Standard Big Bang Cosmology, 
   such as the flatness and horizon problems, 
   but also the primordial density fluctuations which are necessary for the formation 
   of the large scale structure observed in the present universe.  
In a simple inflationary scenario known as the slow-roll inflation, inflation is driven by a single scalar field (inflaton) 
   while inflaton is slowly rolling down its potential to the minimum. 
During the slow-roll, the inflaton potential energy dominates the energy density of the universe, 
   and the universe undergoes an accelerated expansion era, namely, cosmological inflation.   
The inflation ends when the kinetic energy of inflaton starts dominating over its potential energy,  
   and the inflaton eventually decays into particles in the Standard Model (SM).   
The universe is reheated by relativistic particles created from the inflaton decay 
   and continues to the Standard Big Bang Cosmology.

The Planck 2015 results \cite{Planck2015} have set an upper bound on the tensor-to-scalar ratio as $r \lesssim 0.1$,  
    while the best fit value for the spectral index ($n_s$) is $0.9655 \pm 0.0062$ at $68 \%$ CL. 
Hence, the chaotic inflation models with simple inflaton ($\phi$) potentials such as $V \propto  \phi^4$ and $V \propto \phi^2$ 
    are disfavored because of their predictions for $r$ being too large. 
Among many inflation models, quartic inflation with non-minimal gravitational coupling 
    is a very simple model, which can satisfy the constraints from the Planck 2015 results 
    for a non-minimal gravitational coupling $\xi \gtrsim 0.001$ \cite{NonMinimalUpdate}.

In the view point of particle physics, we may think that an inflation model is more compelling 
   if the inflaton also plays an important role in the model. 
The Higgs inflation scenario \cite{Higgs_inflation1,Higgs_inflation2,Higgs_inflation3} 
   is a well-known example, in which the SM Higgs field is identified with the inflaton. 
Also, we may consider a unified scenario between inflaton and dark matter particle \cite{OS}.
When the SM is extended with some extra or unified gauge groups, 
   such extensions always include an extra Higgs field in addition to the SM Higgs field,  
   which is necessary to spontaneously break the gauge symmetry down to the SM one.   
Similarly to the Higgs inflation scenario, we may identify the extra Higgs field with the inflaton.

In this paper, we consider an inflation scenario in the context of the minimal U(1)$_X$ extension 
   of the SM (the minimal U(1)$_X$ model) with the conformal invariance at the classical level \cite{OOT}, 
   where three generations of right-handed neutrinos and a U(1)$_X$ Higgs field 
   are introduced in addition to the SM particle content. 
The minimal U(1)$_X$ model is a generalization of the well-known minimal U(1)$_{B-L}$ model \cite{mBL}, 
   in which the U(1)$_X$ gauge group is realized as 
   a linear combination of the $B-L$ (baryon number minus lepton number) U(1) 
   and the SM U(1)$_Y$ hyper-charge gauge groups \cite{Appelquist:2002mw}.  
The presence of the three right-handed neutrinos is crucial for cancellation of the gauge and mixed-gravitational anomalies, 
   as well as for incorporating the neutrino masses and flavor mixings into the SM via the seesaw mechanism \cite{Seesaw}.

Motivated by the argument in Ref.~\cite{Bardeen:1995kv} that the classical conformal invariance could be a clue 
   for solving the gauge hierarchy problem, we impose the classically conformal invariance on the minimal U(1)$_X$ model. 
Although the conformal invariance is broken at the quantum level, 
   we follow the procedure by Coleman and Weinberg  \cite{CW} 
   and define our model as a massless theory. 
This model possesses interesting properties: The U(1)$_X$ gauge symmetry is radiatively broken 
   via the Coleman-Weinberg mechanism \cite{CW}.  
Associated with this symmetry breaking, the U(1)$_X$ gauge boson ($Z^\prime$ boson) and 
   the right-handed (Majorana) neutrinos acquire their masses. 
Through a mixing quartic coupling between the U(1)$_X$ Higgs and the SM Higgs doublet fields, 
   the electroweak symmetry breaking is triggered once the U(1)$_X$ symmetry is radiatively broken.

In the classically conformal U(1)$_X$  model, we consider the quartic inflation with non-minimal gravitational coupling. 
Here, we identify the U(1)$_X$ Higgs field as the inflaton. 
Because of the symmetry breaking via the Coleman-Weinberg mechanism, 
  the quartic (self-)coupling of the U(1)$_X$ Higgs field relates to the U(1)$_X$ gauge coupling, 
  in other words, we have a relation between the inflaton mass and the $Z^\prime$ boson mass. 
Since the inflationary predictions are controlled by the inflaton quartic coupling
  in the quartic inflation with non-minimal gravitational coupling,  
  we have a correlation between the inflationary predictions and $Z^\prime$ boson physics. 
Assuming the $Z^\prime$ boson mass in the range of ${\cal O}$(10 GeV)$-{\cal O}$(10 TeV), 
  we investigate complementarities between the inflationary predictions and 
  the current constraints from the $Z^\prime$ boson resonance search at the Large Hadron Collider (LHC)
  as well as the prospect of the search for the $Z^\prime$ boson and the right-handed neutrinos 
  at the future collider experiments.

This paper is organized as follows:   
In the next section, we review the basics of the quartic inflation with non-minimal gravitational coupling 
  and the constraints on the inflationary predictions from the Planck 2015 results. 
In Sec.~\ref{sec:3}, we present the classically conformal U(1)$_X$ extended SM, 
  and discuss the interesting property of the model, such as the radiative U(1)$_X$ symmetry breaking 
  and the subsequent electroweak symmetry breaking.     
Identifying the U(1)$_X$ Higgs field as an inflaton, we investigate the quartic inflation 
  with non-minimal  gravitational coupling in Sec.~\ref{sec:4}.  
Because of the radiative U(1)$_X$ symmetry breaking, the inflaton quartic coupling   
  during inflation relates to the U(1)$_X$ gauge coupling at low energies 
  through the renormalization group evolutions. 
In Sec.~\ref{sec:5}, we discuss the current collider constraints on the $Z^\prime$ production cross section 
  and the future prospects of the search for the $Z^\prime$ boson and the right-handed neutrinos. 
Here, we emphasize complementarities between the collider physics and the inflationary predictions. 
For completion of our inflation scenario, we discuss reheating after inflation in Sec.~\ref{sec:6}.  
The last section is devoted to conclusions.

%%%%%%%%%%%%%%%%%%%%%%%%
\section{Non-minimal quartic inflation} 
%%%%%%%%%%%%%%%%%%%%%%%%
\label{sec:2}

In this section, we introduce the quartic inflation with non-minimal gravitational coupling (non-minimal quartic inflation). 
We define the inflation scenario by the following action in the Jordan frame:
\begin{eqnarray}
 {\cal S}_J = \int d^4 x \sqrt{-g} 
   \left[-\frac{1}{2} f(\phi)  {\cal R}+ \frac{1}{2} g^{\mu \nu} \left(\partial_\mu \phi \right) \left(\partial_\nu \phi \right) 
    -V_J (\phi) \right]  , 
\label{S_J}
\end{eqnarray}
where $f(\phi) = (1+ \xi \phi^2)$, $V_J (\phi)$ is the scalar potential and the reduced Planck mass, $M_P=2.44 \times 10^{18}$ GeV, is set to be 1 (Planck unit), 
   $\phi$ is a real scalar (inflaton),  $\xi > 0$ is a dimensionless and real parameter of the non-minimal gravitational coupling,    
    and $\lambda$ is a quartic coupling of the inflaton. 
In the limit $\xi \to 0$, the model is reduced to the minimal quartic inflation.

To obtain an action with a canonically normalized kinetic term for gravity in the so-called Einstein frame, we perform a cannonical transformation of the Jordan frame metric, $ f(\phi) g_{\mu \nu} =  g_{E {\mu \nu}}$, so that 
\begin{eqnarray}
\sqrt{-g} &=& \frac{1}{f(\phi)^2} \sqrt{-g_E}, \nonumber \\
{\mathcal R} &=& f(\phi) \left({\mathcal R}_E -\frac{3}{2} \left(\nabla {\rm ln} f(\phi)\right)^2\right). 
\label{S_E}   
\end{eqnarray}
The action in the Einstein frame is then given by 
\begin{eqnarray}
S_E = \int d^4 x \sqrt{-g_E}\left[-\frac{1}{2}  {\cal R}_E +  \frac{1}{2} \left(\frac{1}{f(\phi)} + \frac{6 \xi^2 \phi^2}{f(\phi)^2} \right)g_E^{\mu \nu} \left(\partial_\mu \phi \right) \left(\partial_\nu \phi \right)
   -\frac{V_J(\phi)}{f(\phi)^2}  \right]. 
\label{S_E}   
\end{eqnarray}
Using a field redefinition,
\begin{eqnarray}
\left(\frac{d\sigma}{d\phi}\right)^{2} = \frac{1+ \xi (6 \xi +1) \phi^2} {\left( 1 + \xi \phi^2 \right)^2}, 
\end{eqnarray}
the scalar kinetic term is canonically normalized and we obtain
\begin{eqnarray}
S_E = \int d^4 x \sqrt{-g_E}\left[-\frac{1}{2}  {\cal R}_E +  \frac{1}{2} g_E^{\mu \nu} \left(\partial_\mu \sigma \right) \left(\partial_\nu \sigma \right)
   -V_E(\phi(\sigma)) \right], 
\label{S_E}   
\end{eqnarray}
where the inflaton potential in the Einstein frame in terms of the original $\phi$ is described as \footnote{Due to the conformal transformation, the SM interaction terms are also scaled by $1/f(\phi)^2$. However, since $\phi \ll 1$ (in Planck units) at the vacuum, the effect of this higher dimensional operator on SM particles is negligible.}
\begin{eqnarray}
V_E =  \frac{\lambda}{4}  \frac{\phi^4}{(1+\xi \phi^2)^2}. 
 \label{VE}
\end{eqnarray}
Note that for large $\phi\gg 1/ \sqrt{\xi}$, $V_E$ becomes a constant. Hence the potential is suitable for the slow-roll inflation.

We express the slow-roll parameters in terms of $\phi$ as follows: 
\begin{eqnarray}
 \epsilon(\phi) &=& \frac{1}{2} \left(\frac{V_E^\prime}{V_E \; \sigma^\prime}\right)^2,   \nonumber \\
 \eta(\phi) &=& \frac{V_E^{\prime \prime}}{V_E \; (\sigma^\prime)^2}- \frac{V_E^\prime \; \sigma^{\prime \prime}}{V_E \; (\sigma^\prime)^3} ,   \nonumber \\
 \zeta (\phi) &=&  \left(\frac{V_E^\prime}{V_E \; \sigma^\prime}\right) 
 \left( \frac{V_E'''}{V_E \; (\sigma^\prime)^3}
-3 \frac{V_E'' \; \sigma''}{V_E \; (\sigma^\prime)^4} 
+ 3 \frac{V_E^\prime \; (\sigma^{\prime \prime})^2}{V_E \; (\sigma^\prime)^5} 
- \frac{V_E^\prime \; \sigma'''}{V_E \; (\sigma')^4} \right)  , 
\end{eqnarray}
where a prime denotes a derivative with respect to $\phi$. 
The amplitude of the curvature perturbation $\Delta_{\cal R}$ is given by 
\begin{equation} 
  \Delta_{\cal R}^2 = \left. \frac{V_E}{24 \pi^2 \epsilon } \right|_{k_0},
\end{equation}
  which should satisfy $\Delta_\mathcal{R}^2= 2.195 \times10^{-9}$
  from the Planck measurements \cite{Planck2015}
  with the pivot scale chosen at $k_0 = 0.002$ Mpc$^{-1}$.
The number of e-folds is given by
\begin{eqnarray}
  N_0 = \frac{1}{\sqrt{2}} \int_{\phi_{\rm e}}^{\phi_0}
  d \phi  \frac{\sigma^\prime}{\sqrt{\epsilon(\phi)}}
\end{eqnarray} 
where $\phi_0$ is the inflaton value at horizon exit of the scale corresponding to $k_0$, 
  and $\phi_e$ is the inflaton value at the end of inflation, 
  which is defined by $\epsilon(\phi_e)=1$.
The value of $N_0$ depends logarithmically on the energy scale during inflation 
  as well as on the reheating temperature, and we take its typical value to be $N_0=50-60$ 
  in order to solve the horizon and flatness problems.

%%%%%%%%%%%%%%%%%%%%%%%%%%%%%%%%%%%%%%%%%%%%%%%%%%%%%
% Fig
%%%%%%%%%%%%%%%%%%%%%%%%%%%%%%%%%%%%%%%%%%%%%%%%%%%%%
\begin{figure}[t]
  \begin{center}
   \includegraphics[scale=1.3]{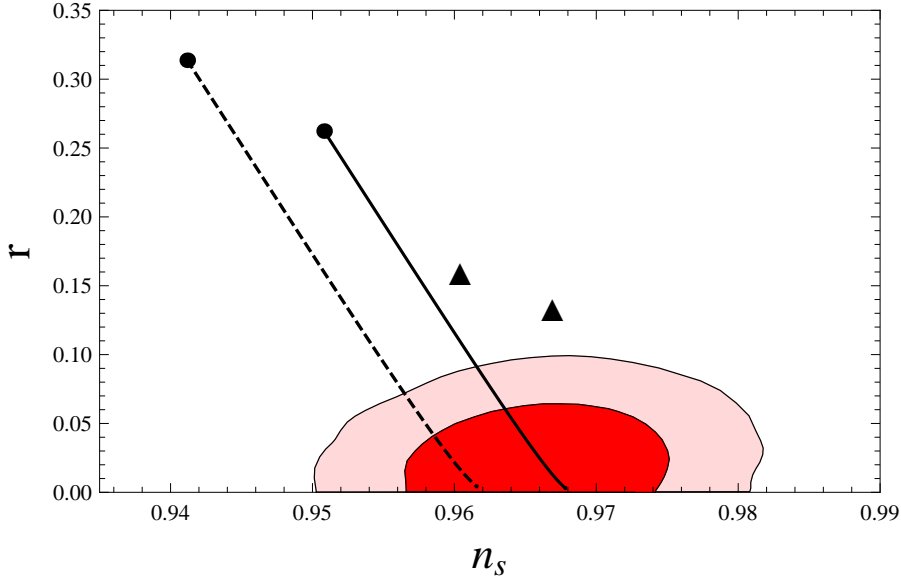}   
   \end{center}
\caption{
The inflationary predictions ($n_s$ and $r$) in the non-minimal quartic inflation for various values of $\xi \geq 0$,   
  along with the contours for the limits at the confidence levels of $68\%$ (inner) and $95\%$ (outer) 
  obtained by the Planck measurements ({\it Planck} TT+lowP+BKP)~\cite{Planck2015}.   
The solid and the dashed diagonal lines correspond to the inflationary predictions
   for $N_0=60$ and $N_0=50$, respectively.  
The predictions of the minimal quartic inflation ($ \xi=0$) for $N_0=60$ and $N_0=50$ 
  are depicted by the right and left black points, respectively.   
Here, we also show the predictions of the quadratic inflation for $N_0=60$ and $N_0=50$   
  as the right and left triangles, respectively. 
As $\xi$ is increased, the predicted $r$ values approach their asymptotic values 
  $r \simeq 0.00296$ and $0.00419$ for $N_0=60$ and $N_0=50$, respectively. 
}
 \label{fig:1}
\end{figure}
%%%%%%%%%%%%%%%%%%%%%%%%%%%%%%%%%%%%%%%%%%%%%%%%%%%%%

%%%%%%%%%%%%%%%%%%%%%%%%%%%%%%%%%%%%%%%%%%%%
\begin{table}[ht]
\begin{center}
\begin{tabular}{|c||cc|ccc|c|}
\hline
\multicolumn{7}{|c|}{$N_0=60$}   \\
\hline 
 $\xi $ &  $\phi_0$ & $\phi_e $ & $n_s$   &  $r$  &  $\alpha (10^{-4})$  & $\lambda$\\ 
\hline
  $0$             & $22.1$  & $2.83$    &  $0.951$ &  $ 0.262$  & $-8.06 $    &  $ 1.43  \times 10^{-13}  $    \\
  $0.00333$  & $22.00$  & $2.79$    & $0.961$  &  $ 0.1$     & $-7.03$      &  $ 3.79  \times 10^{-13}  $    \\
  $0.0689$    & $18.9$    & $2.30$    & $0.967$  &  $ 0.01$   & $-5.44$        &  $ 6.69  \times 10^{-12}  $    \\
  $1$             & $8.52$      & $1.00$       & $0.968$  &  $0.00346$    & $-5.25$ &  $ 4.62  \times 10^{-10}  $    \\
  $10$           & $2.89$      & $0.337$   & $0.968$  &  $0.00301$  & $-5.24$     &  $ 4.01  \times 10^{-8}    $    \\
  $100$         & $0.920$      & $0.107$ & $0.968$  &  $0.00297$ & $-5.23$      &  $ 3.95  \times 10^{-6}    $    \\
  $1000$     & $0.291$ & $0.0340$ & $0.968$  &  $0.00296$ & $-5.23$      &  $ 3.94  \times 10^{-4}    $    \\
\hline
\hline
\multicolumn{7}{|c|}{$N_0=50$}   \\
\hline
 $\xi $ &  $\phi_0$ & $\phi_e $ & $n_s$   &  $r$  &  $\alpha (10^{-4})$  & $\lambda$\\ 
\hline
  $0$             & $20.2$  & $2.83$    &  $0.941$ &  $ 0.314$  & $-11.5 $    &  $ 2.45  \times 10^{-13}  $    \\
  $0.00527$  & $20.0$  & $2.77$    & $0.955$  &  $ 0.1$     & $-9.74$      &  $ 7.83  \times 10^{-13}  $    \\
  $0.119$    & $15.8$    & $2.07$    & $0.961$  &  $ 0.01$   & $-7.70$        &  $ 1.96  \times 10^{-11}  $    \\
  $1$             & $7.82$      & $1.00$       & $0.961$  &  $0.00489$    & $-7.51$ &  $ 6.56  \times 10^{-10}  $    \\
  $10$           & $2.65$      & $0.337$   & $0.962$  &  $0.00426$  & $-7.49$     &  $ 5.70  \times 10^{-8}    $    \\
  $100$         & $0.844$      & $0.107$ & $0.962$  &  $0.00420$ & $-7.48$      &  $ 5.61  \times 10^{-6}    $    \\
  $1000$     & $0.267$ & $0.0340$ & $0.962$  &  $0.00419$ & $-7.48$      &  $ 5.60  \times 10^{-4}    $    \\
\hline
\end{tabular}
\end{center}
\caption{ 
Inflationary predictions for various values of $\xi$ in the non-minimal quartic inflation 
  for fixed $N_0=60$ and $50$. 
Here, $\phi_0$ and $\phi_e$ are evaluated in the Planck units ($M_P=1$).   
} 
\label{Tab:1}
\end{table}
%%%%%%%%%%%%%%%%%%%%%%%%%%%%%%%%%%%%%%%%%%%%%%%%%%%%%%

The slow-roll approximation is valid as long as the conditions $\epsilon \ll 1$, $|\eta| \ll 1$ and $\zeta \ll 1$ hold. 
In this case, the inflationary predictions, 
   the scalar spectral index $n_{s}$, the tensor-to-scalar ratio $r$, 
   and the running of the spectral index $\alpha=\frac{d n_{s}}{d \ln k}$, are given by
\begin{eqnarray}
n_s = 1-6\epsilon+2\eta, \; \;   
r = 16 \epsilon,  \; \;  
\alpha=16 \epsilon \eta - 24 \epsilon^2 - 2 \zeta. 
\end{eqnarray} 
Here, the inflationary predictions are evaluated at $\phi=\phi_0$. 
Under the constraint of $\Delta_\mathcal{R}^2= 2.195\times10^{-9}$ from the Planck measurements \cite{Planck2015}, 
   once $N_0$ is fixed, all the inflationary predictions as well as the quartic coupling $\lambda$  
   are determined as a function of $\xi$. 
In Fig.~\ref{fig:1}, we show the inflationary predictions ($n_s$ and $r$) for various values of $\xi \geq 0$, 
  along with the contours for the limits at the confidence levels of $68\%$ (inner) and $95\%$ (outer) 
  obtained by the Planck measurements ({\it Planck} TT+lowP+BKP)~\cite{Planck2015}.   
The solid and the dashed diagonal lines correspond to the inflationary predictions 
   for $N_0=60$ and $N_0=50$, respectively.  
The predictions of the minimal quartic inflation ($ \xi=0$) for $N_0=60$ and $N_0=50$ 
  are depicted by the right and left black points, respectively.   
Here, we also show the predictions of the quadratic inflation for $N_0=60$ and $N_0=50$   
  as the right and left triangles, respectively. 
As $\xi$ is increased, the inflationary predictions approach their asymptotic values,  
   $n_s \simeq 0.968$,  $r \simeq 0.00296$ and $\alpha \simeq -5.23 \times 10^{-4}$ 
   for $N_0=60$
   ($n_s \simeq 0.962$,  $r \simeq 0.00419$ and $\alpha \simeq -7.48 \times 10^{-4}$ for $N_0=50$). 
In Fig.~\ref{fig:1},  we find a lower bound on $\xi \geq 0.00385$, which corresponds to $r \leq 0.0913$ for $N_0=60$,  
    from the limit at 95\% confidence level. 
We have summarized  in Table~\ref{Tab:1}  
    the numerical values of the inflationary predictions 
    for various $\xi$ values and fixed $N_0=60$ and $50$.

%%%%%%%%%%%%%%%%%%%%%%%%
\section{Classically conformal U(1)$_X$ extended Standard Model } 
%%%%%%%%%%%%%%%%%%%%%%%%
\label{sec:3}

%%%%%%%%%%%%%%%%%%%%%%%%%%%%%%%%%%%%%%
\begin{table}[t]
\begin{center}
\begin{tabular}{|c|ccc|c|}
\hline
      &  SU(3)$_c$  & SU(2)$_L$ & U(1)$_Y$ & U(1)$_X$  \\ 
\hline
$q^{i}_{L}$ & {\bf 3 }    &  {\bf 2}         & $ 1/6$       & $(1/6) x_{H} + (1/3) x_{\Phi}$   \\
$u^{i}_{R}$ & {\bf 3 }    &  {\bf 1}         & $ 2/3$       & $(2/3) x_{H} + (1/3) x_{\Phi}$   \\
$d^{i}_{R}$ & {\bf 3 }    &  {\bf 1}         & $-1/3$       & $(-1/3) x_{H} + (1/3) x_{\Phi}$  \\
\hline
$\ell^{i}_{L}$ & {\bf 1 }    &  {\bf 2}         & $-1/2$       & $(-1/2) x_{H} +(-1) x_{\Phi}$    \\
$e^{i}_{R}$    & {\bf 1 }    &  {\bf 1}         & $-1$                   & $(-1) x_{H} +(-1) x_{\Phi}$   \\
\hline
$H$            & {\bf 1 }    &  {\bf 2}         & $- 1/2$       & $(-1/2) x_{H}$   \\  
\hline
$N^{i}_{R}$    & {\bf 1 }    &  {\bf 1}         &$0$                    & $(-1) x_{\Phi}$     \\
$\Phi$            & {\bf 1 }       &  {\bf 1}       &$ 0$                  & $ (+ 2) x_{\Phi}$  \\ 
\hline
\end{tabular}
\end{center}
\caption{
The particle content of the minimal U(1)$_X$ extended SM. 
In addition to the SM particle content ($i=1,2,3$), the three right-handed neutrinos  
  ($N_R^i$ ($i=1, 2, 3$)) and the U(1)$_X$ Higgs field ($\Phi$) are introduced.   
The U(1)$_X$ charge of a field is determined by two real parameters, $x_H$ and $x_\Phi$, 
  as  $Q_X =  Y x_H + Q_{BL}  \; x_\Phi$ with its hyper-charge ($Y$) and $B-L$ charge ($Q_{BL}$).
Without loss of generality, we fix $x_\Phi=1$ throughout this paper. 
}
\label{table2}
\end{table}
%%%%%%%%%%%%%%%%%%%%%%%%%%%%%%%%%%%%%%%%%%%%%%%

The model we will investigate is the minimal U(1)$_X$ extension of the SM 
  with classically conformal invariance \cite{OOT}, 
  which is based on the gauge group SU(3)$_c \times$SU(2)$_L \times$U(1)$_Y \times$U(1)$_X$. 
The particle content of the model is  listed in Table~\ref{table2}.
In addition to the SM particle content, three generations of right-hand neutrinos (RHNs) $N_R^i$ 
  and a U(1)$_X$ Higgs field $\Phi$ are introduced. 
In the following, the real part of the scalar $\Phi$ is identified with the inflaton.  
The U(1)$_X$ gauge group is defined as a linear combination of 
   the SM U(1)$_Y$  and the U(1)$_{B-L}$ gauge groups, 
   and hence the U(1)$_X$ charges of fields are determined by 
   two real parameters, $x_H$ and $x_\Phi$.   
Since  the charge $x_\Phi$ always appears as a product with the U(1)$_X$ gauge coupling, 
   it is not an independent free parameter of the model, and hence we fix $x_\Phi=1$ throughout this paper. 
We reproduce the minimal $B-L$ model as the limit of $x_H \to 0$. 
The limit of $x_H \to +\infty~(-\infty)$ indicates that the U(1)$_X$ is (anti-)aligned to the SM U(1)$_Y$ direction.  
The anomaly structure of the model is the same as the minimal $B-L$ model \cite{mBL}, and 
  all the gauge and mixed-gravitational anomalies are cancelled in the presence of  
  the three RHNs.   
The covariant derivative relevant to the U(1)$_Y \times$ U(1)$_X$ gauge interaction is given by 
%%%%%%
\begin{equation}
D_\mu = \partial_\mu  
      - i ( g_1 Y + \tilde{g} Q_X ) B_\mu - i g_X  Q_X  Z^{\prime}_\mu, 
 \label{Eq:covariant_derivative}
\end{equation}
%%%%%%%
  where in addition to the U(1)$_Y$ gauge coupling ($g_1$) and the U(1)$_X$ gauge coupling ($g_X$),  
  a new gauge coupling $\tilde{g}$ is introduced from a kinetic mixing between the two U(1) gauge bosons. 
For simplicity, we set $\tilde{g}=0$ at the U(1)$_X$ symmetry breaking scale.  
Although non-zero $\tilde{g}$ is generated in its renormalization group evolution toward high energies, 
   we find that its effect on our final results is negligible.

The Yukawa sector of the SM is extended to have 
\bea
\mathcal{L}_{Yukawa} \supset  - \sum_{i=1}^{3} \sum_{j=1}^{3} Y^{ij}_{D} \overline{\ell^i_{L}} H N_R^j 
          -\frac{1}{2} \sum_{k=1}^{3} Y_M^k \Phi \overline{N_R^{k~C}} N_R^k 
       + {\rm h.c.} ,
\label{Lag1} 
\eea
where the first and the second terms are the neutrino Dirac Yukawa couplings 
  and the Majorana Yukawa couplings, respectively. 
Without loss of generality, the Majorana Yukawa couplings are already diagonalized in our basis.  
Once the U(1)$_X$ Higgs field $\Phi$ develops non-zero vacuum expectation value (VEV), 
   the U(1)$_X$ gauge symmetry is broken and the Majorana masses for the RHNs are generated. 
Then, the light neutrino masses are generated via the seesaw mechanism~\cite{Seesaw} 
   after the electroweak symmetry breaking.  
In this paper, we consider the degenerate mass spectrum for the RHNs, $Y_M^1=Y_M^2=Y_M^3 \equiv Y_M$, for simplicity.

Since we impose the classically conformal invariance on the minimal U(1)$_X$ model,  
   the renormalizable scalar potential at the tree level is given by 
\bea  
V = \lambda_H \left(  H^{\dagger}H  \right)^2
+ \lambda_{\Phi} \left(  \Phi^{\dagger} \Phi   \right)^2
- \lambda_{\rm mix} 
\left(  H^{\dagger}H   \right) 
\left(  \Phi^{\dagger} \Phi  \right) , 
\label{Higgs_Potential}
\eea
where all quartic couplings are chosen to be positive. 
Note that the mass terms for the SM Higgs doublet ($H$) and the U(1)$_X$ Higgs ($\Phi$) 
  are forbidden by the conformal invariance. 
In the following, we assume that $\lambda_{\rm mix}$ is negligibly small (this will be justified later), 
  and analyze the Higgs potential separately for $\Phi$ and $H$ as a good approximation.

Let us first analyze the U(1)$_X$ Higgs sector. 
At the one-loop level, the Coleman-Weinbeg potential \cite{CW} is calculated to be 
\begin{eqnarray}
   V(\phi) =  \frac{\lambda_\Phi}{4} \phi^4 
     + \frac{\beta_\Phi}{8} \phi^4 \left(  \ln \left[ \frac{\phi^2}{v_\phi^2} \right] - \frac{25}{6} \right), 
\label{eq:CW_potential} 
\end{eqnarray}
where $\phi / \sqrt{2} = \Re[\Phi]$ is a real scalar, and 
  we have chosen the renormalization scale as the VEV of $\Phi$ ($\langle \phi \rangle =v_\phi$).  
The stationary condition $\left. dV/d\phi\right|_{\phi=v_\phi} = 0$ leads to a relation,  
\begin{eqnarray}
   \lambda_\Phi = \frac{11}{6} \beta_\Phi, 
\label{eq:stationary}
\end{eqnarray} 
between the renormalized self-coupling defined as 
\begin{eqnarray}
 \lambda_\Phi = \frac{1}{3 !}\left. \frac{d^4V(\phi)}{d \phi^4} \right|_{\phi=v_\phi} 
\end{eqnarray}  
and the coefficient of the one-loop corrections \footnote{In a more precise formulation of the Coleman-Weinberg effective potential, $\beta_\Phi$ includes a $\lambda_{\rm mix}$ term
which we have neglected because it is negligibly small compared to the dominant contribution from $g_X^4$.
Also, we define our inflaton trajectory along the $\phi$ direction with $H = 0$. Hence, even for $\lambda_{\rm mix} \gg \lambda_\Phi$,
we can neglect the $\lambda_{\rm mix}$ term in our inflationary analysis.
},  
\begin{eqnarray}
 \beta_\Phi	= \frac{1}{16 \pi^2}  \left( 20\lambda_\Phi^2 + 96 g_X^4  - 3 Y_M^4 \right) 
	\simeq  
	\frac{1}{16 \pi^2}  \left( 96 g_X^4  - 3 Y_M^4 \right) .  
\end{eqnarray}
Here,  we have used $\lambda_\Phi^2 \ll g_X^4$ in the last expression. 
Note that the U(1)$_X$ symmetry breaking via the Coleman-Weinberg mechanism 
   relates the U(1)$_X$ Higgs quartic coupling to the gauge and Majorana Yukawa couplings 
   in Eq.~(\ref{eq:stationary}). 
The vacuum stability requires $Y_M < (32)^{1/4} g_X$.

We next consider the SM Higgs sector. 
In our model, the electroweak symmetry breaking is achieved in a very simple way. 
Once the U(1)$_X$ symmetry is radiatively broken, 
  the SM Higgs doublet mass is generated through the mixing quartic term 
  in Eq.~(\ref{Higgs_Potential}): 
\begin{equation}
  V \supset  \frac{\lambda_H}{4}h^4 - \frac{\lambda_{\rm mix}}{4} v_\phi^2 h^2,  
\end{equation}
where we have replaced $H$ by $H = 1/\sqrt{2}\, (0 \; \,h)^T$ in the unitary gauge. 
As a result, the electroweak symmetry is broken. 
Here, we emphasize a crucial difference from the SM, 
  namely, the electroweak symmetry breaking is triggered 
  by the radiative U(1)$_X$ gauge symmetry breaking  \cite{Iso:2009ss}, 
  not by a negative mass squared added by hand. 
The SM Higgs boson mass ($m_h$) is given by 
\begin{equation}
  m_h^2  = \lambda_{\rm mix} v_\phi^2 = 2 \lambda_H v_h^2, 
\end{equation}
 where $v_h=246$ GeV is the SM Higgs VEV. 
Considering the Higgs boson mass of $m_h=125$ GeV \cite{mh} and 
  the LEP constraint on $v_\phi \gtrsim 10$ TeV \cite{LEP:2003aa, Carena:2004xs, Schael:2013ita, Heeck:2014zfa},  
  we find $\lambda_{\rm mix} \lesssim 10^{-4}$ and the smallness of $\lambda_{\rm mix}$ is justified.

Associated with the U(1)$_X$ and the electroweak symmetry breakings,  
  the U(1)$_X$ gauge boson ($Z^\prime$ boson) and the (degenerate) Majorana RHNs acquire their masses as 
\begin{eqnarray}
  m_{Z^\prime}  = \sqrt{(2 g_{X} v_\phi)^2  +  (x_H g_{X} v_h)^2} \simeq  2 g_{X} v_\phi, 
   \;  \;  m_N = \frac{Y_M}{\sqrt 2} v_\phi. 
\label{Eq:mass_Zp_DM}
\end{eqnarray} 
The U(1)$_X$ Higgs boson mass is given by 
\begin{eqnarray}
  m_\phi^2 = \left. \frac{d^2 V}{d\phi^2}\right|_{\phi=v_\phi}  
                   =\beta_\Phi v_\phi^2  \simeq 
  \frac{1}{16 \pi^2} \left( 96 g_X^4 - 3 Y_M^4 \right) v_\phi^2 
  =   \frac{6}{\pi} \alpha_X m_{Z^\prime}^2 
     \left( 1-2 \left( \frac{m_N}{m_{Z^\prime}}\right)^4   \right), 
\label{Eq:mass_phi}
\end{eqnarray}
where $\alpha_X = g_X^2/(4 \pi)$. 
The vacuum stability, in other words, $m_\phi^2 >0$, requires 
  $m_{Z^\prime} >  2^{1/4} m_N$.

%%%%%%%%%%%%%%%%%%%%%%%%%%%%%%%%%%%%%%%
\section{Non-minimal quartic inflation with the U(1)$_X$ Higgs field} 
%%%%%%%%%%%%%%%%%%%%%%%%%%%%%%%%%%%%%%%
\label{sec:4}

Now we identify the U(1)$_X$ Higgs filed with the inflaton in the non-minimal quartic inflation. 
In the original Jordan frame action, we introduce the non-minimal gravitational coupling of 
\bea 
  - \xi  \left(  \Phi^\dagger \Phi \right)  {\cal R}, 
\eea
   which leads to the non-minimal gravitational coupling in Eq.~(\ref{S_J}) for the inflaton/Higgs filed defined as $\phi=\sqrt{2} \Re[\Phi]$. 
The scalar potential in Eq.~(\ref{S_J}) is replaced by the effective potential in Eq.~(\ref{eq:CW_potential}).  
Since the inflaton value $\phi \gg v_\phi$ during inflation, we can neglect the effects of the VEV $v_\phi$ 
   for the non-minimal coupling as well as the inflaton potential. 
In our inflation analysis, we employ the renormalization group (RG) improved effective potential 
   of the form \cite{Sher:1988mj}, 
\bea 
  V(\phi) = \frac{1}{4} \lambda_\Phi(\phi)  \phi^4 ,
\eea 
  where $\lambda(\phi)$ is the solution to the RG equation with identifying the renormalization scale 
  as $\phi$ along the inflation trajectory.

As we have discussed in Sec.~\ref{sec:2}, the inflationary predictions are determined 
  by the parameter $\xi$ of the non-minimal gravitational coupling. 
From the view point of the unitarity arguments \cite{Unitarity} of the non-minimal quartic inflation scenario, 
  we may take $\xi \lesssim 10$ to make our analysis valid. 
This means from Table~\ref{Tab:1} that the inflaton quartic coupling is very small, $\lambda \lesssim 4 \times 10^{-8}$ 
  for $N_0=60$. 
Note that the stationary condition of Eq.~(\ref{eq:stationary}) derived from the Coleman-Weinberg mechanism 
    requires the quartic coupling to be very small. 
Hence, one may consider it natural to realize the non-minimal 
  quartic inflation with a small $\xi$ in the context of our classically conformal model. 
Because of the stationary condition and $\lambda_\Phi \ll 1$, the U(1)$_X$ gauge and the Majorana Yukawa couplings 
   must be very small, $g_X$, $Y_M \ll 1$. 
Thus, the RG evolutions of all couplings in our model are very mild, and we calculate the inflationary predictions 
  with a constant quartic coupling, $\lambda_\Phi(\phi_0)$, evaluated at the inflaton value $\phi=\phi_0$. 
Our results for the inflationary predictions in the non-minimal quartic inflation are presented in Sec.~\ref{sec:2}. 
In the following analysis, we identify $\lambda$ in Sec.~\ref{sec:2} with $\lambda= \lambda_\Phi(\phi_0)$.

We evaluate the inflaton quartic coupling at $\phi=\phi_0$ by extrapolating the gauge, the Majorana Yukawa,  
   and the Higgs quartic couplings at $v_\phi$ through their RG equations. 
Since all couplings are very small, the RG equations at the one-loop level are approximately given by    
\bea 
  \frac{d \lambda_\Phi}{d \ln \phi} &=& \beta_{\lambda} \simeq 
   96 \alpha_X^2 - 3 \alpha_Y^2,  \nonumber \\
  \frac{d \alpha_X}{d \ln \phi } &=& \beta_g = \frac{72 + 64 x_H + 41 x_H^2}{12 \pi}  \alpha_X^2,  \nonumber \\
\frac{d \alpha_Y}{d \ln \phi} &=& \beta_Y = 
    \frac{1}{2 \pi} \alpha_Y 
   \left( \frac{5}{2} \alpha_Y - 6  \alpha_X \right) ,
\label{ApproxRGE}  
\eea
where $\alpha_Y=Y_M^2/(4 \pi)$. 
In the leading-log approximation, we have the solutions of the RG equations for $\alpha_X$ and $\alpha_Y$  as 
\bea 
  \alpha_X(\phi) \simeq  \overline{\alpha_X} + \overline{\beta_g} \ln \left[\frac{\phi}{v_\phi} \right], \; \; 
  \alpha_Y(\phi) \simeq  \overline{\alpha_Y} + \overline{\beta_Y} \ln \left[\frac{\phi}{v_\phi} \right], 
\eea 
where  $\overline{\alpha_X} \equiv \alpha_X(v_\phi)$, $\overline{\alpha_Y} \equiv \alpha_Y(v_\phi)$, 
   and $\overline{\beta_g}$ and $\overline{\beta_Y}$ are the beta functions in Eq.~(\ref{ApproxRGE})  
   evaluated with $\overline{\alpha_X}$ and $\overline{\alpha_Y}$. 
Using these solutions, we obtain 
\bea
 \beta_\lambda \simeq 96 \alpha_X^2 - 3 \alpha_Y^2   
 \simeq \overline{\beta_\lambda} + 2 \left( 96 \; \overline{\alpha_X} \; \overline{\beta_g} -3  \; \overline{\alpha_Y} \; \overline{\beta_Y} 
 \right) \ln \left[\frac{\phi}{v_\phi} \right] ,  
\eea
where $\overline{\beta_\lambda}=96 \; \overline{\alpha_X}  -3  \; \overline{\alpha_Y} $. 
Finally, we arrived at an approximate solution as 
\bea 
 \lambda_\Phi(\phi) &\simeq &
   \overline{\lambda_\Phi} + \overline{\beta_\lambda} \ln \left[\frac{\phi}{v_\phi} \right] 
  + \left( 96 \; \overline{\alpha_X} \; \overline{\beta_g} -3  \; \overline{\alpha_Y} \; \overline{\beta_Y} 
 \right) \left( \ln \left[\frac{\phi}{v_\phi} \right] \right)^2  \nonumber \\
 & = &
  \left( \frac{11}{6} + \ln \left[\frac{\phi}{v_\phi} \right] \right) \overline{\beta_\lambda} 
  + \left( 96 \; \overline{\alpha_X} \; \overline{\beta_g} -3  \; \overline{\alpha_Y} \; \overline{\beta_Y} 
 \right) \left( \ln \left[\frac{\phi}{v_\phi} \right] \right)^2  ,  
\label{RGEsol}
\eea
where $ \overline{\lambda_\Phi} \equiv  \lambda_\Phi(v_\phi) $, and we have used Eq.~(\ref{eq:stationary}) in the second line.

\begin{figure}[t]
\begin{center}
\includegraphics[scale=0.9]{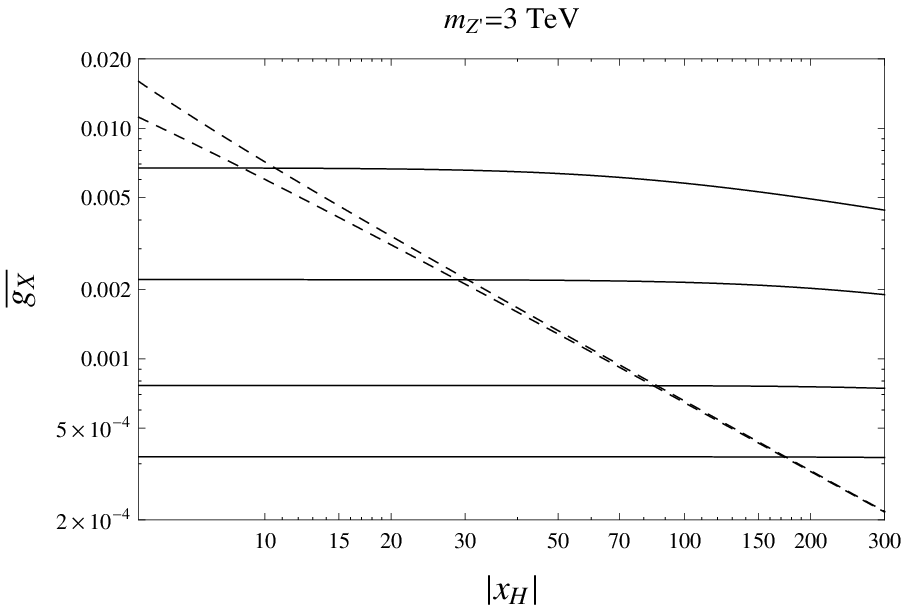} \;
\includegraphics[scale=0.9]{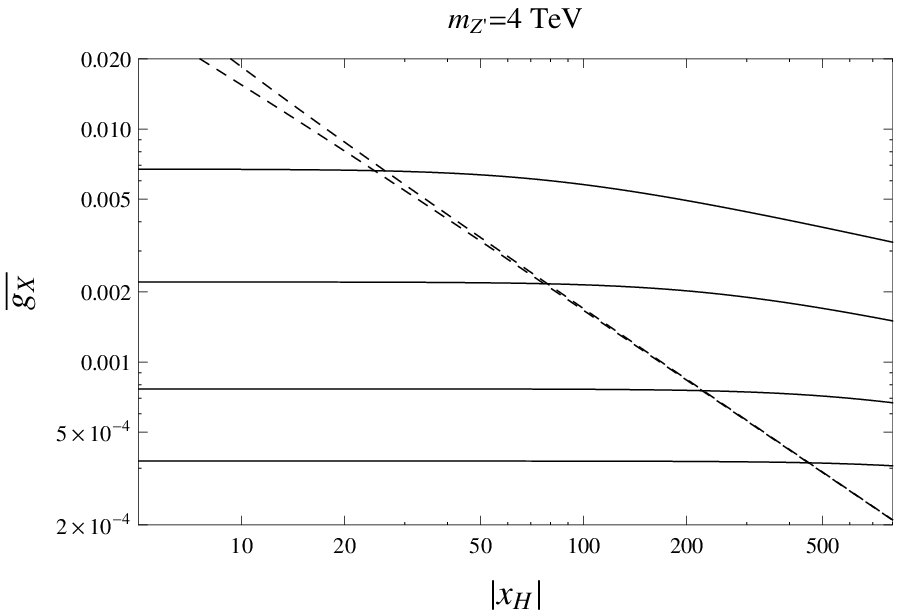}       
\end{center}
\caption{
Left panel: The horizontal solid lines depict the U(1)$_X$ gauge coupling $\overline{g_X}$ as a function of $x_H$ 
  for various values of $\xi=10$, $1$, $0.0689$, and $0.00333$ from top to bottom, 
  along which the non-minimal quartic inflation is realized.  
Here, we have fixed $m_{Z^\prime}=3 \, m_N=3$ TeV.    
The result solid lines for $x_H>0$ and $x_H<0$ are well overlapped and indistinguishable. 
The dashed lines show the upper bounds on $\overline{g_X}$ as a function of $x_H$ from the ATLAS results 
  on the search for a narrow resonance \cite{ATLAS:2017}.   
The upper and lower dashed lines correspond to $x_H < 0$ and $x_H > 0$, respectively. 
Right panel: same as the left panel, but for $m_{Z^\prime}=3 \,m_N=4$ TeV.    
}
\label{fig:2}
\end{figure}

In the next section, we will discuss the collider physics for the $Z^\prime$ boson and the heavy Majorana neutrinos. 
For our discussion, it is convenient to adopt 
  the $Z^\prime$ boson mass ($m_{Z^\prime}$) and the degenerate heavy Majorana neutrino mass ($m_N$) 
  as free parameters, instead of the U(1)$_X$ Higgs VEV $v_\phi$ and $\overline{Y_M}$. 
In our analysis, we have 5 free parameters, namely,  
  $\xi$, $x_H$, $\overline{g_X}$, $m_{Z^\prime}$, and $m_N$, 
  after replacing $v_\phi$ and $\overline{Y_M}$ by using the relations, 
  $v_\phi = m_{Z^\prime}/(2 \, \overline{g_X})$ and 
  $\overline{Y_M}= \sqrt{2} m_N/v_\phi  = 2 \sqrt{2} \; \overline{g_X} \, (m_N/m_{Z^\prime})$.
As has been discussed in Sec.~\ref{sec:2}, once $\xi$ is fixed,  
   not only the inflationary predictions but also $\phi_0$, $\phi_e$ and $\lambda_\Phi(\phi_0)$ are all determined. 
When $\xi$, $m_{Z^\prime}$ and $m_N$ values are fixed, 
   we obtain $\overline{g_X}$ as a function of $x_H$ from Eq.~(\ref{RGEsol}). 
In Fig.~\ref{fig:2}, we show $\overline{g_X}$ as a function of $x_H$ 
   for various values of $\xi$ for $m_{Z^\prime}=3$ TeV (left panel) and 4 TeV (right panel). 
In each panel, the horizontal solid lines correspond to $\xi=10$, $1$, $0.0689$, and $0.00333$ from top to bottom. 
Here, we have fixed $m_N=m_{Z^\prime} /3$ (see the next section), for simplicity.  
The results for $x_H>0$ and $x_H<0$ are well overlapped and indistinguishable. 

%%%%%%%%%%%%%%%%%%%%%%%%%%%%%%%%%
%%%%%%%%%%%%%%%%%%%%%%%%%%%%%%%%%

%%%%%%%%%%%%%%%%%%%%%%%%%%%%%%%%%%%
\section{Complementarity between collider physics and inflation}
%%%%%%%%%%%%%%%%%%%%%%%%%%%%%%%%%%%
\label{sec:5}

Realizing the non-minimal quartic inflation in the context of the classically conformal U(1)$_X$ model, 
  we have obtained a relation between the U(1)$_X$ gauge coupling and the inflationary predictions 
  once $x_H$, $m_{Z^\prime}$ and $m_N$ are fixed. 
If $m_{Z^\prime} \lesssim 10$ TeV, the $Z^\prime$ boson in our U(1)$_X$ model 
  can be produced at the high-energy colliders. 
Since the production cross section of the $Z^\prime$ boson depends on its mass, 
   the gauge coupling and $x_H$, we have in our model a correlation 
   between the collider physics on $Z^\prime$ boson and the inflationary predictions.

Let us first consider the LHC phenomenology on $Z^\prime$ boson. 
The ATLAS and CMS collaborations have been searching for a narrow resonance 
   with dilepton final states at the LHC Run-2 \cite{ATLAS:2016, CMS:2016}.  
In their analysis, the so-called sequential SM $Z^\prime$ ($Z^\prime_{SSM}$) 
   has been considered as a reference, 
   assuming the $Z^\prime_{SSM}$ boson has the exactly the same properties 
   as the SM $Z$ boson, except for its mass.  
In the following, we interpret the current LHC constraints on the $Z^\prime_{SSM}$ boson 
   into the U(1)$_X$ $Z^\prime$ boson to identify an allowed parameter region.  
In our analysis, we employ the latest upper bound on the $Z^\prime_{SSM}$ production cross section 
   reported by the ATLAS collaboration \cite{ATLAS:2017}.

The cross section for the process $pp \to Z^\prime +X \to \ell^{+} \ell^{-} +X$  is given by
\begin{eqnarray}
 \sigma
 =  \sum_{q, {\bar q}}
 \int d M_{\ell \ell} 
 \int^1_ \frac{M_{\ell \ell}^2}{s} dx
 \frac{2 M_{\ell \ell}}{x s}  
 f_q(x, Q^2) f_{\bar q} \left( \frac{M_{\ell \ell}^2}{x s}, Q^2
 \right)  {\hat \sigma} (q \bar{q} \to Z^\prime \to  \ell^+ \ell^-) ,
\label{X_LHC}
\end{eqnarray}
where $M_{\ell \ell}$ is the invariant mass of a final state dilepton,  
  $f_q$ is the parton distribution function for a parton (quark) ``$q$'', 
  and $\sqrt{s} =13$ TeV is the center-of-mass energy of the LHC Run-2.
In our numerical analysis, we employ CTEQ6L~\cite{CTEQ} 
  for the parton distribution functions with the factorization scale $Q= m_{Z^\prime}$. 
The cross section for the colliding partons is given by
\bea 
{\hat \sigma}(q \bar{q} \to Z^\prime \to  \ell^+ \ell^-) =
 \frac{\pi}{1296}  \overline{\alpha_X}^2 
 \frac{M_{\ell \ell}^2}{(M_{\ell \ell}^2-m_{Z^\prime}^2)^2 + m_{Z^\prime}^2 \Gamma_{Z^\prime}^2} 
F_{q \ell}(x_H),  
\label{X_LHC2}
\eea
where the function $F_{q \ell}(x_H)$ are 
\bea
   F_{u \ell}(x_H) &=&  (8 + 20 x_H + 17 x_H^2)  (8 + 12 x_H + 5 x_H^2),   \nonumber \\
   F_{d \ell}(x_H) &=&  (8 - 4 x_H + 5 x_H^2) (8 + 12 x_H + 5 x_H^2) 
\label{Fql}
\eea
for ``$q$'' being the up-type ($u$) and down-type ($d$) quarks, respectively. 
Since the RG running effect from $m_{Z^\prime}$ to $v_\phi$ is negligible, 
  we use $\overline{\alpha_X}=\overline{g_X}^2/(4 \pi)$ for the U(1)$_X$ gauge coupling 
  in our collider physics analysis. 
Neglecting the mass of all SM fermions, the total decay width of $Z^\prime$ boson is given by 
\bea
\Gamma_{Z'} = 
 \frac{\overline{\alpha_X}}{6} m_{Z^\prime} 
 \left[ F(x_H) + 3 \left( 1- \frac{4 m_N^2}{m_{Z^\prime}^2} \right)^{\frac{3}{2}} 
 \theta \left( \frac{m_{Z^\prime}}{m_N} - 2 \right)  \right] 
\label{width}
\eea
with $ F(x_H)=13+ 16 x_H  + 10 x_H^2 $.

In interpreting the latest ATLAS results \cite{ATLAS:2017} 
  on the $Z^\prime_{SSM}$ boson into the U(1)$_X$ $Z^\prime$ boson case, 
  we follow the strategy in Ref.~\cite{OO}: 
   we first calculate the cross section of the process $pp \to Z^\prime_{SSM} +X \to \ell^{+} \ell^{-} +X$, 
   and then we scale our result by a $k$-factor so as to match with the theoretical prediction 
   of the cross section presented in the ATLAS paper   \cite{ATLAS:2017}. 
With the $k$-factor determined in this way, we calculate the cross section 
   for the process $pp \to Z^\prime+X \to \ell^{+} \ell^{-} +X$  
   to identify an allowed region for the model parameters of $\overline{g_X}$, $x_H$ and $m_{Z^\prime}$.

In Fig.~\ref{fig:2}, the dashed lines show the upper bounds on $\overline{g_X}$ as a function 
  of $x_H$ from the ATLAS results on the search for a narrow resonance 
  with the combined dielectron and dimuon channels \cite{ATLAS:2017}. 
The upper and lower dashed lines correspond to $x_H < 0$ and $x_H > 0$, respectively.     
As we can see the cross section formula, the dashed lines approach with each other for a large $|x_H|$.    
Combining the ATLAS constraints with the horizontal lines from the inflationary analysis, 
   we find upper bounds on    
   $x_H \lesssim 10$, $30$, $80$, and $170$ for $m_{Z^\prime}=3$ TeV 
   ($x_H \lesssim 25$, $80$, $220$, and $450$ for $m_{Z^\prime}=4$ TeV), 
   corresponding to $\xi=10$, $1$, $0.0689$, and $0.00333$, respectively.  
Recall that the inflaton quartic coupling is extremely small for $\xi \lesssim 10$ (see Table~\ref{Tab:1}), 
   and this indicates that the U(1)$_X$ gauge coupling is also very small (see Eq.~(\ref{RGEsol})). 
Nevertheless, as has been pointed out in Ref.~\cite{OOR},  
   the $Z^\prime$ boson with mass of ${\cal O}$(1 TeV) can still be tested at the LHC Run-2  
   when the U(1)$_X$ gauge symmetry is oriented to the SM U(1)$_Y$ hyper-charge direction,
   namely, $|x_H| \gg 1$.

%%%%%%%%%%%%%%%%%%%%%%%%%%%%%%%%%%%%%%%%%%%%%
\begin{figure}[t]
\begin{center}
\includegraphics[scale=0.9]{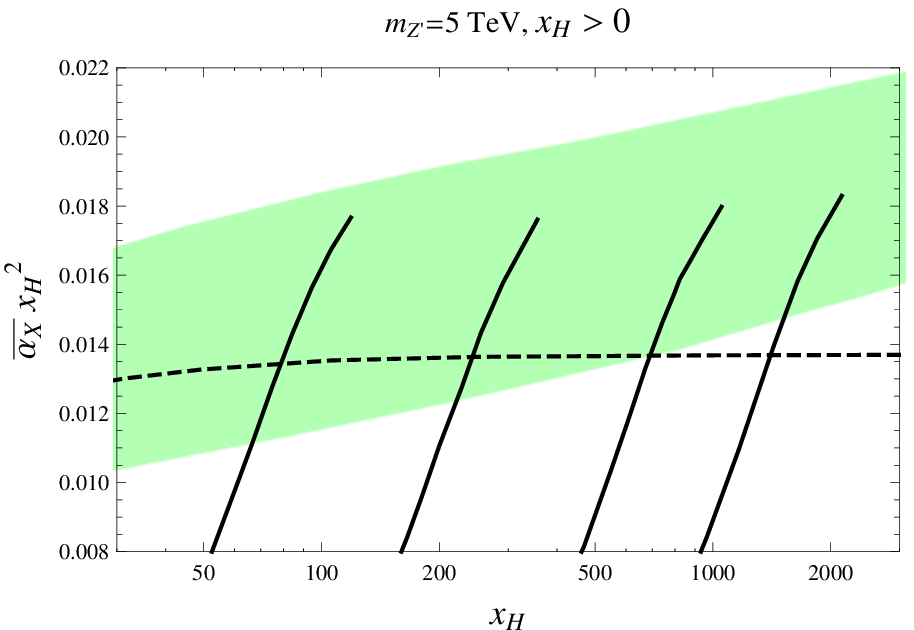} \;
\includegraphics[scale=0.9]{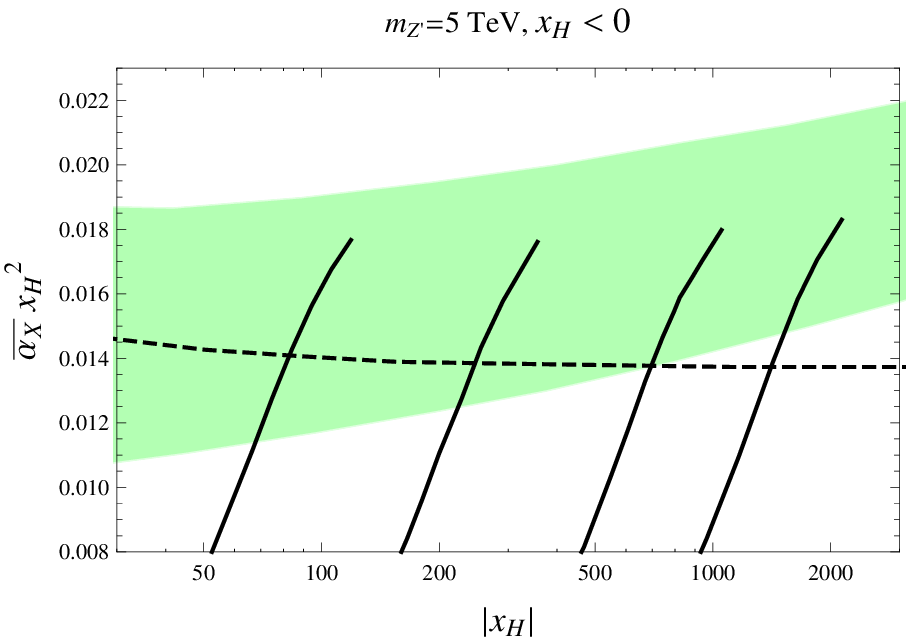}       
\end{center}
\caption{
Left panel: the combined result for $m_{Z^\prime}=5$ TeV and $x_H>0$. 
The shaded (green) region depicts the parameters 
  to resolve the electroweak vacuum instability, 
   while satisfying the perturbativity of the gauge coupling at $M_P$.   
The dashed line denotes the upper bound from the ATLAS results 
   for the $Z^\prime$ boson search at the LHC Run-2.  
The diagonal lines correspond to $\xi=10$, $1$, $0.0689$, and $0.00333$ from left to right, 
   along which the non-minimal quartic inflation is realized.  
Right panel: same as the left panel, but for $x_H < 0$. 
}
\label{fig:3}
\end{figure}
%%%%%%%%%%%%%%%%%%%%%%%%%%%%%%%%%%%%%%%%%%%%%%

As the $Z^\prime$ boson is heavier, the current LHC bounds become weaker, 
   because of the energy dependence of the parton distribution functions.  
We can see this fact by comparing the dashed lines in the left and right panels of Fig.~\ref{fig:2}. 
When we take $m_{Z^\prime}=5$ TeV, which is the maximum $Z^\prime$ boson mass
   in the ATLAS analysis \cite{ATLAS:2017}, 
   another interesting parameter region of our model opens up. 
In Ref.~\cite{OOT},  the same model presented in this paper has been investigated 
  in the view point of the electroweak vacuum stability. 
As is well-known, the SM Higgs potential becomes unstable at high energies, 
  since the running SM Higgs quartic coupling runs into the negative region 
  at the renormalization scale of $\mu \simeq 10^{10}$ GeV \cite{Buttazzo:2013uya}.  
It has been shown in Ref.~\cite{OOT} that this electroweak vacuum instability problem 
  can be solved in the context of the classically conformal U(1)$_X$ model with $\overline{\alpha_X} \, x_H^2 \gtrsim 0.01$. 
It is interesting to combine our inflation analysis with the results in Ref.~\cite{OOT}.

Fig.~\ref{fig:3} shows the combined results in $(x_H, \, \overline{\alpha_X} \, x_H^2)$-plane.  
In the left panel, the parameter region to resolve the electroweak vacuum instability is 
  shown as the shaded (green) region for $m_{Z^\prime}=5$ TeV and $x_H>0$. 
In order to solve the instability problem, $\overline{\alpha_X} \, x_H^2 \gtrsim 0.01$ is necessary, 
  while $\overline{\alpha_X}$ has an upper bound for a fixed $x_H$ 
  from the requirement $\alpha_X(M_P) <1$ that the running U(1)$_X$ gauge coupling is in the perturbative regime at $\mu=M_P$.  
The dashed line denotes the upper bound from the ATLAS results. 
The diagonal lines correspond to $\xi=10$, $1$, $0.0689$, and $0.00333$ from left to right, 
  along which the non-minimal quartic inflation is realized.  
Since we have found that the leading-log approximation for the RG analysis is not sufficiently reliable
   for $\overline{\alpha_X} \, x_H^2 \gtrsim 0.01$, 
   we have numerically integrated the RG equations in this analysis. 
See Ref.~\cite{OOT} for details of our RG analysis. 
The upper bounds on $\overline{\alpha_X} \, x_H^2 \lesssim 0.018$ shown on the diagonal lines 
   are also from the requirement of $\alpha_X(\phi_0) < 1$ for a given $\xi$. 
Since $\phi_0 > M_P$ for $\xi \lesssim 10$, the requirement of $\alpha_X(\phi_0) < 1$ is more severe 
  than that of $\alpha_X(M_P) < 1$.     
We find the allowed parameter region for $\xi \gtrsim 0.0689$ and $x_H \lesssim 700$, 
   although it is very narrow. 
The right panel is the same as the left panel, but for $x_H < 0$.

%%%%%%%%%%%%%%%%%%%%%%%%%%%%%%%%%%%%%%%%%%%%%%%%%%%%%
% Fig
%%%%%%%%%%%%%%%%%%%%%%%%%%%%%%%%%%%%%%%%%%%%%%%%%%%%%
\begin{figure}[t]
  \begin{center}
   \includegraphics[scale=0.45, angle=90]{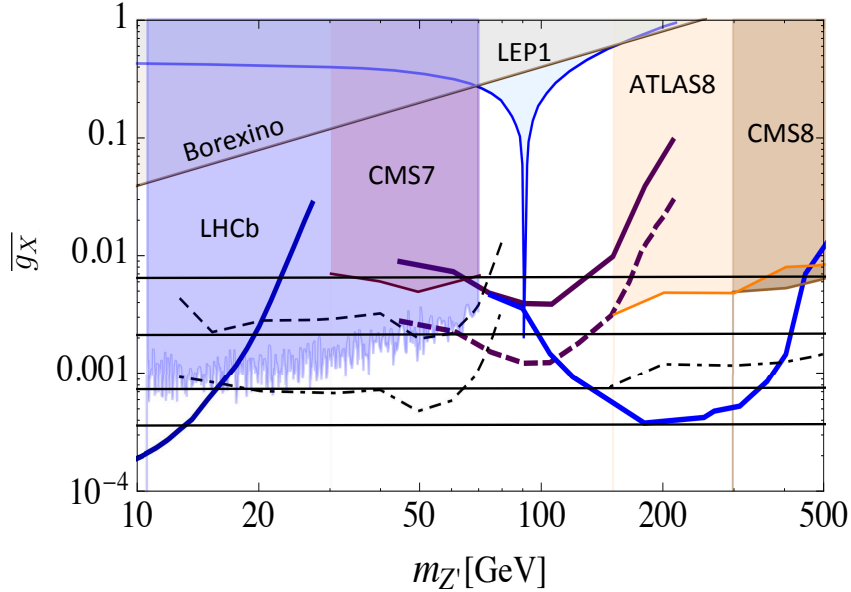}   
   \end{center}
\caption{
The $B-L$ gauge coupling ($\overline{g_X}$) a function of $m_{Z^\prime}$, 
   along with the results presented in Ref.~\cite{Batell:2016zod}. 
We also show the current bound from the LHCb results \cite{LHCb}. 
The horizontal lines correspond to our results for 
   $\xi=10$, $1$, $0.0689$, and $0.00333$ from top to bottom, respectively,  
   along which the non-minimal quartic inflation is realized.  
According to the analysis in Ref.~\cite{Batell:2016zod}, we have fixed $m_N=m_{Z^\prime}/3$. 
The shaded regions are excluded by the indicated experiments. 
The projected reach of the proposed searches for a $Z^\prime$ boson production 
  and its decay into a pair of RHNs are shown in thick (solid and dashed) curves. 
The thin (black) curves show the projected sensitivity of direct searches for the $Z^\prime$ boson production 
  via its decay $Z^\prime \to \ell^+ \ell^-$ from the LHC Run-1 (dashed), and the High-Luminosity LHC (dot-dashed).
See Ref.~\cite{Batell:2016zod} for more details. 
}
 \label{fig:4}
\end{figure}
%%%%%%%%%%%%%%%%%%%%%%%%%%%%%%%%%%%%%%%%%%%%%%%%%%%%%

Even if the U(1)$_X$ gauge coupling is very small and $|x_H| \lesssim 1$,  
   we can test our model when the $Z^\prime$ boson is light, say, $m_{Z^\prime} \lesssim 500$ GeV.  
In Ref.~\cite{Batell:2016zod}, the authors have considered the RHN production 
  at the High-Luminosity LHC \cite{HL-LHC}  and the SHiP \cite{SHiP} experiments 
  in the contest of the minimal $B-L$ model (the limit of $x_H=0$ in our U(1)$_X$ model), 
  where a pair of RHNs is created through the decay of a $Z^\prime$ boson 
  resonantly produced at the colliders. 
When the RHNs have the mass of ${\cal O}$(100 GeV) or less, 
   it is long-lived and its decay to the SM particles provides a clean signature with a displaced vertex.   
It has been found in Ref.~\cite{Batell:2016zod} that for a fixed $m_N =m_{Z^\prime}/3$, 
  the High-Luminosity  LHC and the SHiP experiments can explore the $B-L$ gauge coupling 
  up to $\overline{g_X} \gtrsim 10^{-4}$ for $10 \; {\rm GeV} \lesssim m_{Z^\prime}  \lesssim 500$ GeV.  
In the $B-L$ limit of $x_H=0$,  
  we show in Fig.~\ref{fig:4} the $B-L$ gauge coupling ($\overline{g_X}$) as a function of $m_{Z^\prime}$, 
  along with the results presented in Ref.~\cite{Batell:2016zod}. 
In Fig.~\ref{fig:4}, we have added the current bound from the LHCb results \cite{LHCb}. 
The horizontal lines correspond to our results for 
   $\xi=10$, $1$, $0.0689$, and $0.00333$ from top to bottom, respectively,  
   along which the non-minimal quartic inflation is realized.  
Our results very weakly depend on $m_{Z^\prime}$ in the mass range shown in Fig.~\ref{fig:4}, 
   as can be understood from Eq.~(\ref{RGEsol}).

%%%%%%%%%%%%%%%%%%%%%%%%%%%%%%%%%%%%%%%%%%%%%%%%%%%%%
% Fig
%%%%%%%%%%%%%%%%%%%%%%%%%%%%%%%%%%%%%%%%%%%%%%%%%%%%%
\begin{figure}[t]
  \begin{center}
   \includegraphics[scale=1.2]{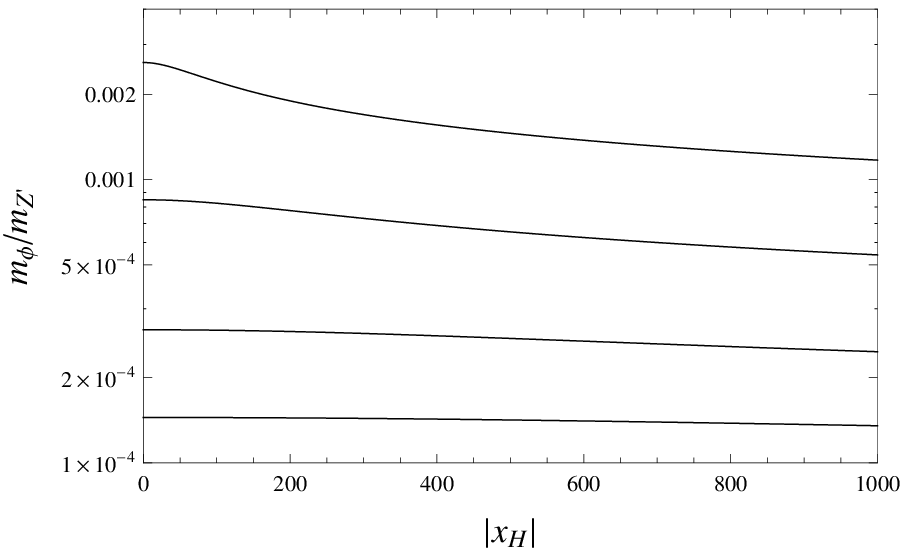}   
   \end{center}
\caption{
The mass ratio of $m_\phi/m_{Z^\prime}$ as a function of $x_H$  
   for $\xi=10$, $1$, $0.0689$, and $0.00333$ from top to bottom. 
Although we have used $m_{Z^\prime}=3$ TeV as a reference,  
  we obtain almost identical results for other values of  $m_{Z^\prime}$. 
}
 \label{fig:5}
\end{figure}
%%%%%%%%%%%%%%%%%%%%%%%%%%%%%%%%%%%%%%%%%%%%%%%%%%%%%
  
%%%%%%%%%%%%%%%%%%%%%%%%%%%%%%%%%%%
\section{Inflaton mass and reheating after inflation}
%%%%%%%%%%%%%%%%%%%%%%%%%%%%%%%%%%%
\label{sec:6}

To complete our inflation scenario, we finally discuss reheating after inflation 
   through the inflaton decay into the SM particles. 
Since the inflaton is much lighter than the $Z^\prime$ boson and the RHNs in our scenario with $\xi \lesssim 10$, 
   it decays mainly into the SM fermions through the mixing with the SM Higgs boson.

From the Higgs potential in Eq.~(\ref{Higgs_Potential}) with the radiative corrections in Eq.~(\ref{eq:CW_potential}), 
  we find the following mass matrix for the inflaton ($\phi$) and the SM Higgs boson ($h$) 
  at the potential minimum: 
\bea
{\cal L}  \supset -
\frac{1}{2}\begin{bmatrix}h  & \phi\end{bmatrix}
\begin{bmatrix} 
m_h^2 &  - m_{\rm mix}^2 \\ 
- m_{\rm mix}^2 & m_{\phi}^2
\end{bmatrix} 
\begin{bmatrix} h \\ \phi \end{bmatrix}, 
\label{HiggsMassMatrix}
\eea 
where $m_{\rm mix}^2= \lambda_{\rm mix} v_h v_\phi$,  $m_h=125$ GeV and $m_\phi$ is given in Eq.~(\ref{Eq:mass_phi}). 
As can be seen in Sec.~\ref{sec:3}, 
  $m_{\rm mix}^2$, $m_\phi^2 \ll m_h^2$ and the mass matrix is almost diagonal. 
We define the mass eigenstates, $\phi_1$  and $\phi_2$, by  
\begin{eqnarray}
\begin{bmatrix} h \\ \phi \end{bmatrix}   =
\begin{bmatrix} \cos\theta &   -\sin\theta \\ \sin\theta & \cos\theta  \end{bmatrix} \begin{bmatrix} \phi_1 \\ \phi_2 
\end{bmatrix}  ,
\end{eqnarray} 
with a small mixing angle 
\bea
   \theta \simeq \frac{m_{\rm mix}^2}{m_h^2}  =2 \overline{g_X} \, \left(\frac{v_h}{m_{Z^\prime}} \right) \ll 1.
\label{angle}
\eea
Since the mixing angle is very small, the mass eigenstate $\phi_1$ ($\phi_2$) is almost the SM Higgs boson (the U(1)$_X$ Higgs boson).

Through the mixing angle, the inflaton decays into the SM particles. 
We evaluate the inflaton decay width as 
\bea 
   \Gamma_{\phi} \simeq \theta^2 \times  \Gamma_h(m_{\phi}) , 
\label{width_phi}   
\eea
where $\Gamma_h(m_{\phi})$ is the SM Higgs boson decay width 
   if the SM Higgs boson mass were $m_{\phi}$. 
From Eqs.~(\ref{Eq:mass_phi}) and (\ref{angle}),  
  the inflaton mass and its decay width is a function of  $\overline{\alpha_X}$ and $m_{Z^\prime}$ 
  (with $m_N=m_{Z^\prime}/3$). 
For the successful non-minimal inflation,  $\overline{\alpha_X}$ is determined 
  as a function of $\xi$, $x_H$ and $m_{Z^\prime}$, and 
  hence the inflaton mass and the decay width are controlled by the three parameters, 
  $\xi$, $x_H$ and $m_{Z^\prime}$. 
With the inflaton decay width, we estimate reheating temperature by 
\bea 
  T_{\rm RH} = \left( \frac{90}{\pi^2 g_*} \right)^{1/4} \sqrt{\Gamma_\phi M_P} \simeq \sqrt{\Gamma_\phi M_P}, 
\eea  
where $g_*$ is the total effective degrees of freedom of thermal plasma.

In Fig.~\ref{fig:5}, we show the ratio of $m_\phi/m_{Z^\prime}$ as a function of $x_H$  
   for $\xi=10$, $1$, $0.0689$, and $0.00333$ from top to bottom. 
The results for $x_H > 0$ and $x_H < 0$ are well overlapped and indistinguishable.      
Although we have used $m_{Z^\prime}=3$ TeV as a reference, we find that 
  the result is almost independent of $m_{Z^\prime}$, as we have seen in Fig.~\ref{fig:4} 
  with $x_H=0$.  
The resultant mass ratios are also weakly depending on $x_H$.      

%%%%%%%%%%%%%%%%%%%%%%%%%%%%%%%%%%%%%%%%%%%%%
\begin{figure}[t]
\begin{center}
\includegraphics[scale=0.9]{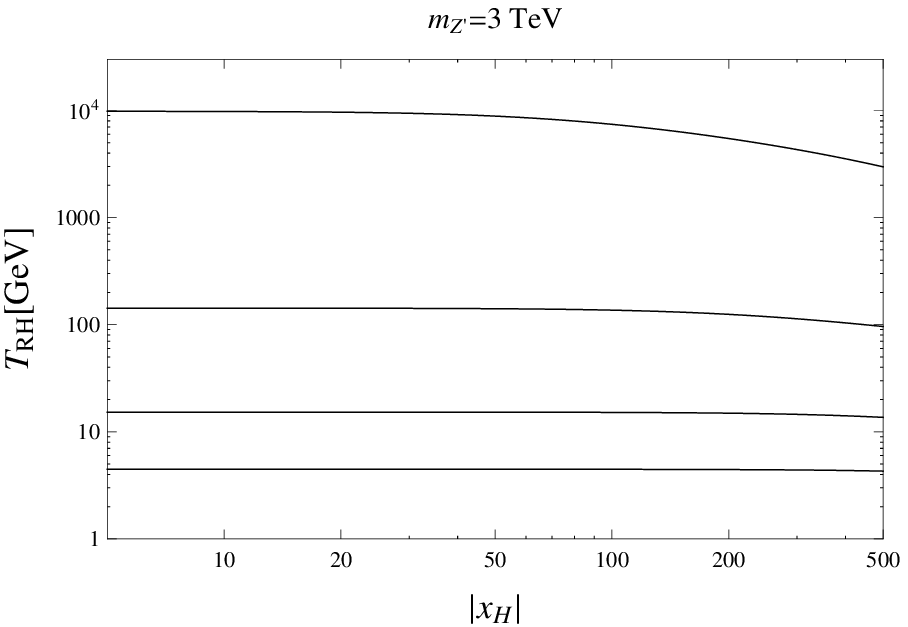} \;
\includegraphics[scale=0.9]{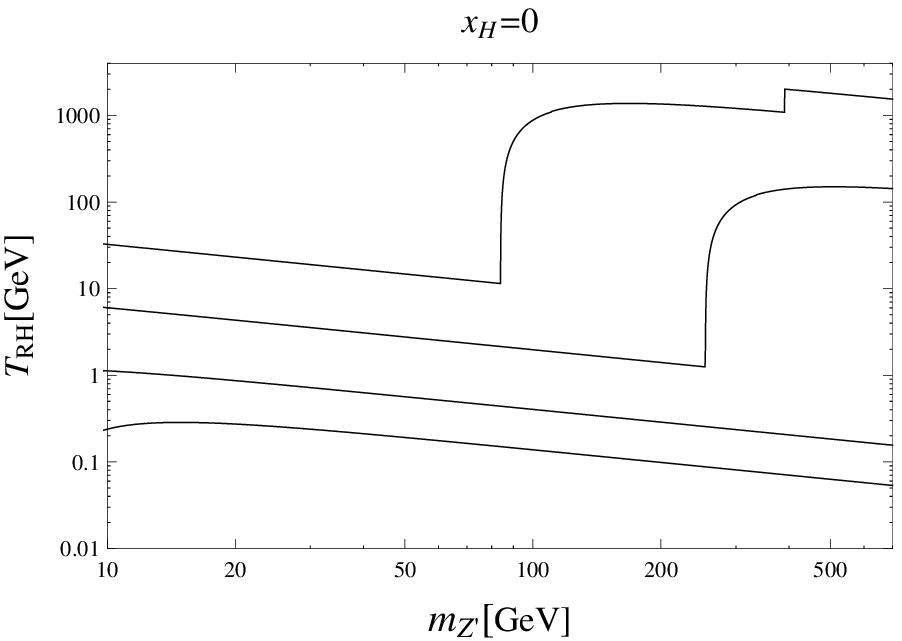}       
\end{center}
\caption{
Reheating temperature after inflation. 
Left panel: reheating temperature as a function of $x_H$  
   for $\xi=10$, $1$, $0.0689$, and $0.00333$ from top to bottom, with $m_{Z^\prime}=3$ TeV. 
The results for $x_H>0$ and $x_H<0$ are well overlapped and indistinguishable.   
Right panel: reheating temperature as a function of $m_{Z^\prime}$ in the $B-L$ model ($x_H=0$).    
The solid lines correspond to the results for $\xi=10$, $1$, $0.0689$, and $0.00333$ from top to bottom. 
Sharp rises of the reheating temperature for threshold values of $m_{Z^\prime}$  imply 
  that new decay channels are opened. 
%For example, in the plot for $\xi=10$, a new decay channel of $\phi \to \mu^+ \mu^-$ 
%  opens at $m_{Z^\prime} \simeq 80$ GeV.  
%A model-independent lower bound on the reheating temperature $T_{\rm RH} \gtrsim 1$ MeV 
%  is satisfied in all the results presented here.
}
\label{fig:6}
\end{figure}
%%%%%%%%%%%%%%%%%%%%%%%%%%%%%%%%%%%%%%%%%%%%%%

In Fig.~\ref{fig:6}, we show the estimated reheating temperature after inflation. 
The left panel depicts the reheating temperature as a function of $x_H$  
   for $\xi=10$, $1$, $0.0689$, and $0.00333$ from top to bottom, with $m_{Z^\prime}=3$ TeV. 
For the $B-L$ limit of $x_H=0$, the right panel depicts the results as a function of $m_{Z^\prime}$. 
The solid lines from top to bottom correspond to the results for $\xi=10$, $1$, $0.0689$, and $0.00333$, respectively. 
Sharp rises of the reheating temperature for threshold values of $m_{Z^\prime}$  imply 
  that new decay channels are opened. 
For example, in the plot for $\xi=10$, a new decay channel of $\phi \to \mu^+ \mu^-$ 
  opens at $m_{Z^\prime} \simeq 80$ GeV.  
All results presented in Fig.~\ref{fig:6} satisfy the model-independent lower bound 
  on reheating temperature, $T_{\rm RH} \gtrsim 1$ MeV, for the successful Big Bang Nucleosynthesis.

%%%%%%%%%%%%%%%%%%%%%%%%%%%%%%%%%%
\section{Conclusions}
%%%%%%%%%%%%%%%%%%%%%%%%%%%%%%%%%%
\label{sec:7}

The non-minimal quartic inflation is a simple and successful inflation scenario, 
   and its inflationary predictions are consistent with the Planck 2015 results 
   for the non-minimal gravitational coupling with $\xi \gtrsim 0.003$ for $N_0=60$. 
This inflation scenario would be more compelling if the inflaton plays essential roles 
   for not only inflation but also particle physics phenomena.  
In many models beyond the SM where the gauge symmetry of the SM is extended, 
    a new Higgs field to break the extended gauge symmetry is commonly introduced. 
It is an interesting possibility to identify such a Higgs field with the inflaton 
   in the non-minimal quartic inflation.

In this paper, we have considered the classically conformal U(1)$_X$ extended SM, 
   where the U(1)$_X$ gauge group is realized as a linear combination of 
   the U(1)$_{B-L}$ and the SM U(1)$_Y$ gauge groups.  
This model has an interesting property that all the gauge symmetry breakings 
   in the model originates from the Coleman-Weinberg mechanism: 
The U(1)$_X$ gauge symmetry is radiatively broken through the Coleman-Weinberg mechanism, 
   and this breaking generates a negative mass squared for the SM Higgs doublet 
   and hence, the electroweak symmetry breaking occurs subsequently.  
Associated with the U(1)$_X$ gauge symmetry breaking, 
   the $Z^\prime$ boson and the right-handed neutrinos acquire their masses. 
We have set their masses in the range of ${\cal O}$(10 GeV)$-{\cal O}$(10 TeV),  
   which is accessible at high energy collider experiments

We have investigated the non-minimal inflation scenario in the context of this classically conformal U(1)$_X$ model 
   by identifying the U(1)$_X$ Higgs field with the inflaton. 
In this model, the U(1)$_X$ gauge symmetry is radiatively broken through the Coleman-Weinberg mechanism, 
   due to which the inflaton quartic coupling is determined by the U(1)$_X$ gauge coupling. 
Since the inflationary predictions in the non-minimal quartic inflation are determined 
   by the inflaton quartic coupling during inflation,
   we have a correlation between the inflationary predictions and the U(1)$_X$ gauge coupling. 
With this correlation, we have investigated complementarities between the inflationary predictions and 
  the current constraint from the $Z^\prime$ boson resonance search at the LHC Run-2  
  as well as the prospect of the search for the $Z^\prime$ boson and the right-handed neutrinos 
  at the future collider experiments. 
For completion of our inflation scenario, we have considered a reheating scenario 
  due to the inflaton decay through the SM Higgs boson, and found the reheating temperature 
  to be sufficiently high.

Here, we comment on the stability of the scalar potential during inflation. 
We have considered the inflation trajectory in the direction of $\phi$  with $H=0$. 
For $\phi \gg v_\phi$, the scalar potential is approximated by Eq.~(\ref{Higgs_Potential}) 
  with replacing the quartic couplings at the tree-level by their RG running couplings. 
If $\lambda_{\rm mix} > 0$ during inflation, we can see a problem 
  that the inflaton potential is destabilized in the SM Higgs direction. 
In Ref.~\cite{OOT}, the authors have shown the numerical result of RG evolution   of $\lambda_{\rm mix}$ from the 1 TeV scale to Planck scale, from which we can see that the $\lambda_{\rm mix}$ quickly changes its sign around 1 TeV in the RG evolution. 
We can easily see this behavior from the RG equation for $\lambda_{\rm mix}$  at 1-loop level, which is approximately given by \cite{OOT}
\bea
\phi \frac{d \lambda_{\rm mix}}{d \phi} \simeq -\frac{1}{16\pi^2}\; 12 x_H^2 g_X^4,
\eea
 for $|x_H|\gg1$. Since the beta function is negative and its absolute value is greater that initial value of $\lambda_{\rm mix}$ at the TeV scale, we can see that $\lambda_{\rm mix}$ quickly becomes negative in its running. 
Although the beta function formula becomes very complicated (see \cite{OOT} for complete formulas) for a small $|x_H|$ value, we obtain the same consequence.

In our analysis we have considered the number of e-folds to be a free parameter and have fixed $N_0= 60$. However, the number of e-folds is determined by the reheating temperature $T_R$, and the inflaton potential energy  at the horizon exit ($V_E \left. \right|_{k_0}$) as (see, for example, Ref.~\cite{Lyth})
\bea
N_0 \simeq 51.4 +\frac{2}{3} {\rm ln} \left(\frac{V_E \left. \right|_{k_0}^{1/4}}{10^{15} {\rm GeV}}\right) +\frac{1}{3} {\rm ln}\left(\frac{T_R}{10^{7} {\rm GeV}}\right). 
\label{efolds}
\eea
Because of this relation, the number of e-folds is not a free parameter and is determined as a function of $\xi, x_H$, and $m_{Z^\prime}$. 
Using this relation we can make our predictions more precise. 
However, in such an analyis the inflationary predictions, low energy observables, and the reheating temperature are related with each other in a very complicated way through the free parameters $\xi, x_H$, and $m_{Z^\prime}$. 
To keep our discussion very clear we have treated $N_0$ as a free parameter. 
From Eq.~(\ref{efolds}), we can see that the true value of $N_0$ lies in between 50 and 60. 
As shown in Table.~\ref{Tab:1}, the inflationary predictions for a fixed $\xi$ weakly depend on $N_0$ values. 
Hence our results with $N_0 = 60$ well approximate the true values.

%%%%%%%%%%%%%%%%%%%%%%%%%%%%%%%%%%
\section*{Acknowledgments}
%%%%%%%%%%%%%%%%%%%%%%%%%%%%%%%%%%
The work of S.O. and D.-s.T. is supported by 
Mathematical and Theoretical Physics unit [Hikami unit] and Advanced Medical Instrumentation unit [Sugawara unit], 
respectively, of the Okinawa Institute of Science and Technology Graduate University.
The work of N.O. is supported in part by the United States Department of Energy (DE-SC0012447). 

%%%%%%%%%%%%%%
{}
%%%%%%%%%%%%%%%%%%%%%%%%%%%%%%%%%%%%%%%%%%%%%%%%

\end{document}